\documentclass[11pt]{article}
\usepackage{graphicx}
\usepackage{epsfig}

\newcommand{\BABARPubYear}    {00}
\newcommand{\BABARProcNumber} {28}
\newcommand{\SLACPubNumber} {8690}

\def\lbabar{\mbox{{\large\sl B}\hspace{-0.4em} {\normalsize\sl A}\hspace{-0.03em}{\large\sl B}\hspace{-0.4em} {\normalsize\sl A\hspace{-0.02em}R}}}
\usepackage{relsize}
\def\babar{\mbox{\slshape B\kern-0.1em{\smaller A}\kern-0.1em
    B\kern-0.1em{\smaller A\kern-0.2em R}}}

\def\epem       {\ensuremath{e^+e^-}}

\def\Kbar  {\kern 0.2em\overline{\kern -0.2em K}{}}

\def\Kstarz  {\ensuremath{K^{*0}}}

\def\Kzb   {\ensuremath{\Kbar^0}}
\def\KzKzb {\ensuremath{K^0 \kern -0.16em \Kzb}}

\def\Dbar  {\kern 0.2em\overline{\kern -0.2em D}{}}

\def\Dzb   {\ensuremath{\Dbar^0}}
\def\DzDzb {\ensuremath{D^0 {\kern -0.16em \Dzb}}}
\def\Dstar   {\ensuremath{D^*}}

\def\Bz    {\ensuremath{B^0}}
\def\B     {\ensuremath{B}}
\def\Bbar  {\kern 0.18em\overline{\kern -0.18em B}{}}

\def\Bzb   {\ensuremath{\Bbar^0}}

\def\BB    {\ensuremath{B\Bbar}} 
\def\BzBzb {\ensuremath{B^0 {\kern -0.16em \Bzb}}}

\def\jpsi  {\ensuremath{{J\mskip -3mu/\mskip -2mu\psi\mskip 2mu}}} 
\mathchardef\Upsilon="7107
\def\Y#1S{\ensuremath{\Upsilon{(#1S)}}}

\def\FourS {\Y4S}

\mathchardef\Deltares="7101
\mathchardef\Xi="7104
\mathchardef\Lambda="7103
\mathchardef\Sigma="7106
\mathchardef\Omega="710A
\def\Deltabar   {\kern 0.25em\overline{\kern -0.25em \Deltares}{}}
\def\Lbar {\kern 0.2em\overline{\kern -0.2em\Lambda\kern 0.05em}\kern-0.05em{}}
\def\Sigbar{\kern 0.2em\overline{\kern -0.2em \Sigma}{}}
\def\Xibar{\kern 0.2em\overline{\kern -0.2em \Xi}{}}
\def\Obar{\kern 0.2em\overline{\kern -0.2em \Omega}{}}
\def\Nbar{\kern 0.2em\overline{\kern -0.2em N}{}}
\def\Xbar{\kern 0.2em\overline{\kern -0.2em X}{}}

\def\mes        {\mbox{$m_{\rm ES}$}}

\def\ev   {\ensuremath{\rm \,e\kern -0.08em V}}
\def\kev  {\ensuremath{\rm \,ke\kern -0.08em V}} 
\def\mev  {\ensuremath{\rm \,Me\kern -0.08em V}} 
\def\gev  {\ensuremath{\rm \,Ge\kern -0.08em V}} 
\def\gevc {\ensuremath{{\rm \,Ge\kern -0.08em V\!/}c}} 
\def\tev  {\ensuremath{\rm \,Te\kern -0.08em V}}
\def\mevc {\ensuremath{{\rm \,Me\kern -0.08em V\!/}c}} 
\def\gevcc{\ensuremath{{\rm \,Ge\kern -0.08em V\!/}c^2}} 
\def\mevcc{\ensuremath{{\rm \,Me\kern -0.08em V\!/}c^2}}

\def\mum  {\ensuremath{\,\mu\rm m}} 

\def\invfb   {\ensuremath{\mbox{\,fb}^{-1}}}
\def\mus  {\ensuremath{\rm \,\mus}}

\def\ps   {\ensuremath{\rm \,ps}}

\def\mus        {\ensuremath{\,\mu{\rm s}}}    
\def\ps         {\ensuremath{{\rm \,ps}}}   
\def\gsim{{~\raise.15em\hbox{$>$}\kern-.85em
          \lower.35em\hbox{$\sim$}~}}
\def\lsim{{~\raise.15em\hbox{$<$}\kern-.85em
          \lower.35em\hbox{$\sim$}~}}

\def\CP                 {\ensuremath{C\!P}}

\def\to                 {\ensuremath{\rightarrow}}

\def\pep2{PEP-II}

\def\mistag{\ensuremath{w}}

\def\deltaz{\ensuremath{{\rm \Delta}z}}
\def\deltat{\ensuremath{{\rm \Delta}t}}
\def\deltamd{\ensuremath{{\rm \Delta}m_d}}

\providecommand{\eqref}[1]{Eq.~(\ref{eq:#1})}

\newcommand{\epjc}      [1]  {{Eur.\ Phys.\ Jour.\ C~{\bf #1}}}

\def\jetset74   {\mbox{\tt Jetset \hspace{-0.5em}7.\hspace{-0.2em}4}}

\providecommand{\xpm}{\mbox{$\pm$}}

\providecommand{\btodstarlnu}{\mbox{$B\to D^{*}l\nu$}}

\providecommand{\B}{\mbox{$B$}}

\providecommand{\bztodstarlnu}{\mbox{$B^0\to D^{*-}\ell^+\nu$}}

\long\def\inst#1{\par\nobreak\kern 4pt\nobreak
    {\it #1}\par\vskip 10pt plus 3pt minus 3pt}

\begin{document}
{\pagestyle{empty}

\begin{flushright}
SLAC-PUB-\SLACPubNumber \\
\babar-PROC-\BABARPubYear/\BABARProcNumber \\
November 2000 \\
\end{flushright}

\par\vskip 4cm

\begin{center}
\Large \bf A study of \boldmath \BzBzb\ oscillations 
with full reconstructed \B\ mesons with the \babar\ detector 
\end{center}
\bigskip

\begin{center}
\large 
Shahram Rahatlou\\
University of California, San Diego  \\
Physics Department, 9500 Gilman Drive, La Jolla CA 94306, USA \\
representing the \lbabar\ Collaboration
\end{center}
\bigskip \bigskip

\begin{center}
\large \bf Abstract
\end{center}

\noindent
Time--dependent \BzBzb\ flavor oscillations are studied in
\epem\ annihilation data collected with the
\babar\ detector at center-of-mass energies near the \FourS\ resonance.
We report a preliminary result for the  time-dependent \BzBzb\ oscillation frequency,
$\deltamd = 0.512 \xpm\  0.017 \xpm\ 0.022$~$\hbar$\ps$^{-1}$.

\vfill
\begin{center}
Contribued to \\
The Meeting of \\
The Division of Particle and Fields \\
of the American Physical Society \\
Columbus, Ohio, USA \\
August 9 - August 12, 2000
\end{center}

\vspace{1.0cm}
\begin{center}
{\em Stanford Linear Accelerator Center, Stanford University, 
Stanford, CA 94309} \\ 
\vspace{0.1cm}\hrule\vspace{0.1cm}
Work supported in part by Department of Energy contract DE-AC03-76SF00515.
\end{center}

\newpage

\newcommand{\secname}{}

\setcounter{footnote}{0}

\renewcommand{\secname}{Introduction}
\section{Introduction}

\noindent
We have performed a measurement of time-dependent
mixing at the PEP-II asymmetric $e^+e^-$ collider at SLAC, where
resonant production of the \FourS\  provides a copious source of \BzBzb\
pairs. The data set used for this analysis corresponds to an integrated luminosity
of 8.9\invfb\ on the \FourS\ resonance and 0.8\invfb\ collected 40\mev\
below the resonance. This corresponds to about $10.1 \times 10^6$
produced \BB\ pairs.

The \babar\ detector is described in detail elsewhere
\cite{BabarPub0017}.  The analysis described here uses
all the detector capabilities, including high resolution
tracking and calorimetry, particle identification and vertexing.


\renewcommand{\secname}{Event Reconstruction}
\section{Event Reconstruction}

\noindent
We fully reconstruct one  \B\ meson~($\B_{\rm REC}$) in hadronic
(\Bz $\to D^{(*)-} \pi^+$, $D^{(*)-} \rho^+$,
$D^{(*)-} a_1^+$ and $\jpsi \Kstarz$) or 
semileptonic ($\bztodstarlnu$) decay mode \footnote{Throughout this paper, 
charge conjugate modes are implied.}. A total of 2577 neutral \B\ candidates is
reconstructed in hadronic decay modes, with an average purity close to $90\%$. 
The main background for these modes is combinatorial. 7517 \Bz\ candidates
are reconstructed in the semileptonic mode, with an average purity close to
$84\%$. Backgrounds to the semileptonic mode are due to combinatorial \Dstar\, 
fake leptons, uncorrelated \Dstar\ $l$ combinations, $c\bar c$ events, 
and charged \B\ decays from $B^-\rightarrow D^{*+}(n\pi) l^- \nu$. 

The other two important ingredients for this analysis are the vertex
reconstruction
and the identification of the flavor of the other \B\ meson~($\B_{\rm
TAG}$) in the event. The flavor of the $\B_{\rm TAG}$  is
determined from the correlation between the particle types and the
charge of its decay products \cite{BabarPub0008}. 
If there is an identified lepton its charge
is used; otherwise the summed charge of identified kaons provides the tag.
An event with no tagging leptons or kaons can still be tagged by the
use of a neural network that exploits the flavor information carried by
other decay products, such as soft leptons from charm semileptonic decays and soft pions
from \Dstar\ decays. 

At \pep2\, the \B\ meson pairs produced in the decay of the
\FourS\ resonance are moving in the lab frame along the
beam axis ($z$ direction) with a Lorentz boost of $\beta_z \gamma = 0.56$.
The separation between the two \B\ vertices along the boost direction,
$\Delta z = z_{\rm REC} - z_{\rm TAG}$, is measured and used to
estimate the decay time difference, $\Delta t \approx \Delta z/\beta_z
\gamma c$.
The  $\B_{\rm TAG}$ vertex is determined via
an inclusive procedure applied to all tracks 
not associated with the $\B_{\rm REC}$ 
meson \cite{BabarPub0001}. 
The typical separation between the two vertices is
$\deltaz = \beta_z \gamma c \tau_B \approx 260$\mum, to be compared to the
experimental resolution $\sim 100$\mum.
The $\Delta t$ resolution is limited by the precision on the
$\B_{\rm TAG}$ vertex, and has little dependence on the decay mode of
the $\B_{\rm REC}$. The $\Delta t$ resolution function is well described
by three Gaussians: core, tail and outlier. We calculate the uncertainty
on $\Delta t$ by using a globally-fitted rescaling of the event-by-event 
vertex separation errors. Most of the events, $\sim 70\%$, are in the 
 core Gaussian, with $\sigma \sim 0.6$ ps.

\renewcommand{\secname}{Likelihood fit method}
\section{Likelihood Fit method}

\noindent
The time-dependent asymmetry between same sign \Bz\Bz $/$\Bzb\Bzb\ (unmixed) and
 opposite sign \Bz\Bzb\ (mixed) events,
$A(\Delta t) = (N_{unmix}-N_{mix})/(N_{unmix}+N_{mix})$
is calculated as a function of $\Delta t$ and is given by

 $$A(\Delta t) \approx (1-2\mistag) \cos \deltamd
\deltat \otimes {\cal {R}}( \Delta t | \hat{a} ) ,$$  

\noindent
where $\hat a$ are the parameters of the $\Delta t$ resolution function
\cite{BabarPub0008} and \mistag\ is the probability of incorrect tagging
(mistag fraction). A simultaneous unbinned  likelihood fit to the
$\Delta t$  distribution of mixed and unmixed events in all  tagging categories,
assuming a common resolution function, allows the simultaneous determination of both 
\deltamd\ and the mistag fractions, $\mistag_i$.
An empirical description of the $\Delta t$ structure of the backgrounds
is determined from a fit to background control samples taken from data, 
allowing for the following components: zero lifetime, non-zero lifetime with no
mixing, non-zero lifetime with mixing.

\renewcommand{\secname}{Results and Conclusions}
\section{Results and Conclusions}

\noindent
We measure the \BzBzb\ oscillation frequency to be
$\Delta m_d  =  0.516 \pm 0.031\ ({\rm stat})  \pm 0.018  ({\rm
  syst})\  \hbar {\rm ps}^{-1}$ in the hadronic sample and  
$\Delta m_d    =   0.508 \pm 0.020\ ({\rm stat}) {}\pm 0.022 ({\rm
    syst})\ \hbar  {\rm ps}^{-1}$ in the $D^{*-}\ell^+\nu$ sample .
 Figure \ref{fig:mixing} shows the asymmetry
  $A(\Delta t)$ distributions for each sample with the fit result superimposed.

The systematic errors include uncertainty due to Monte
Carlo statistics, $\Delta t$ resolution function, background $\Delta
t$ shape,  fraction of background events, \Bz\ lifetime, $z$ scale and 
the boost. In addition, we have looked at the uncertainty due to
feeddown from $B^-\rightarrow D^{*+}(n\pi) l^- \nu$ in the
semileptonic sample. The dominant contribution in the hadronic sample
comes from the $\Delta t$ resolution function, while the semileptonic
sample is dominated by the uncertainty on the fraction of background
events (see \cite{BabarPub0008} for details).

\begin{figure}[htb]
\vspace{9pt}
\begin{center}
\begin{tabular}{cc}   
  \epsfig{file=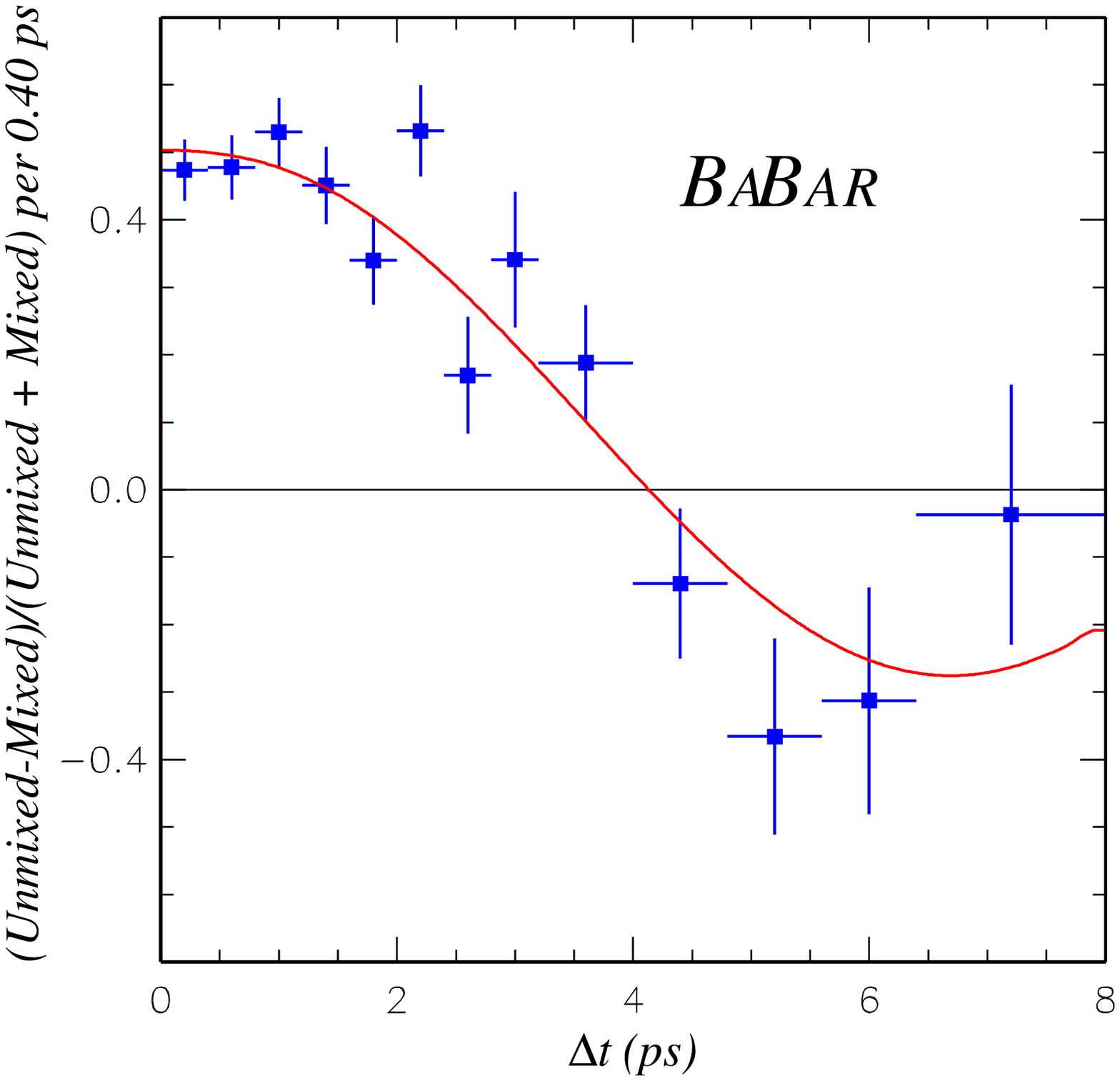,height=5.0cm}  &
  \epsfig{file=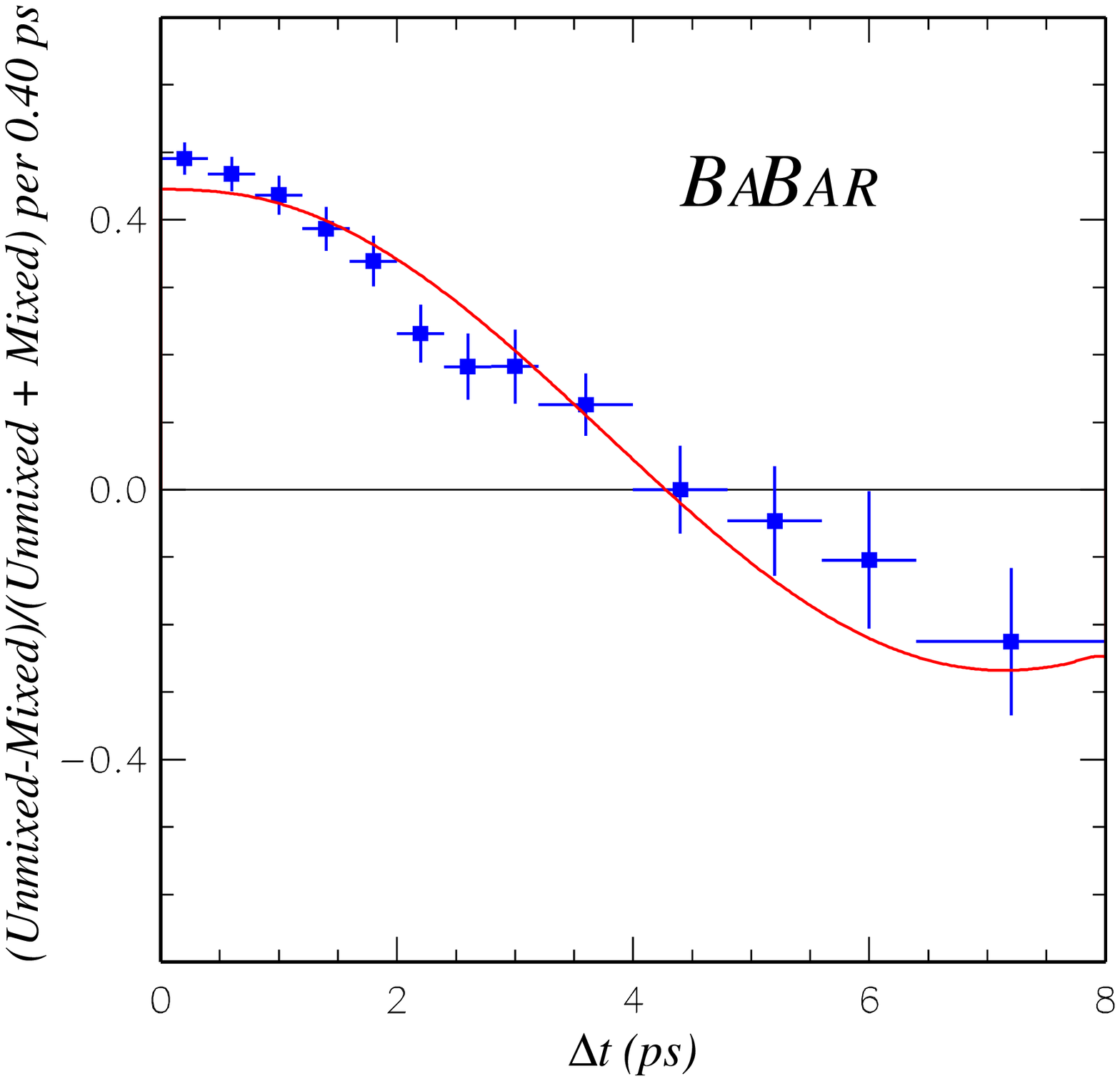,height=5.0cm}  \\
\end{tabular}
\caption{Time-dependent asymmetry $A(\Delta t)$ between unmixed and
      mixed events for (left) hadronic $B$ candidates with $\mes >
      5.27$\gevcc and (right) for \btodstarlnu\ candidates.}
\label{fig:mixing}
\end{center}
\end{figure}

\noindent
Combining the two \deltamd\ results, we obtain the preliminary result: 
\begin{eqnarray*}
\Delta m_d &=& 0.512 \pm  0.017 ({\rm stat})  \pm 0.022 ({\rm syst})\ \hbar {\rm ps}^{-1}.
\end{eqnarray*}

The effective flavor tagging efficiency is given by
$Q=\sum_i \epsilon_i (1-2\mistag_i)^2$ where 
the sum is over tagging categories, each characterized by 
a tagging efficiency $\epsilon_i$ and a mistag fraction $\mistag_i$.  
$Q$ is related to the statistical significance of the measurement 
($1/\sigma_{stat}^2 \sim N_{\B_{\rm TAG}} Q$) and is found to be
$(27.9 \pm 1.6)\%$. The mistag fractions and the $\Delta t$ resolution
function parameters are used in the \CP\ asymmetry measurement \cite{BabarPub0001}.

The results for $\Delta m_d$ are consistent with previous
measurements~\cite{PDG} and  are of
similar precision. They are also compatible with 
other \babar\ measurements \cite{christophe1,christophe2}. Significant
improvements are expected in the near future with the accumulation of
more data and further systematic studies.






\begin{thebibliography}{99}

\bibitem{BabarPub0017}
\babar\ Collaboration, 
B.\ Aubert {\em et al.},
SLAC-PUB-8540, BABAR-CONF-00/01,
contributed to ICHEP2000


\bibitem{BabarPub0008}
\babar\ Collaboration, 
B.\ Aubert {\em et al.},
Preprint hep-ex/0008052.


\bibitem{BabarPub0001} 
\babar\ Collaboration, 
B.\ Aubert {\em et al.},
Preprint hep-ex/0008048.



\bibitem{PDG} D.E. Groom, {\em et al.}, \epjc{15}(2000) 1.


\bibitem{christophe1} 
\babar\ Collaboration, 
B.\ Aubert {\em et al.},
Preprint hep-ex/0008053.


\bibitem{christophe2} 
\babar\ Collaboration, 
B.\ Aubert {\em et al.},
Preprint hep-ex/0008054.





\end{thebibliography}
\end{document}